\documentclass[preprintnumbers,amsmath,amssymb,prl,floatfix,twocolumn,aps]{revtex4}
\usepackage{graphicx}
\usepackage{ulem}

\begin{document}

\title{Topologically-protected metallic states induced by a one-dimensional 
extended defect in a 2D topological insulator}

\author{E. N. Lima}

\affiliation{Departamento de F\'{\i}sica, ICEx, Universidade Federal de Minas
Gerais, Avenida Ant\^onio Carlos 6627, Pampulha, 30123-970, Belo
Horizonte, MG, Brazil}

\altaffiliation{Current addres: Departamento de Matem\'atica, ICEN, Universidade Federal de Mato Grosso,
Rodovia Rondon\'opolis-Guiratinga, KM 06 (MT-270), Sagrada Fam\'{\i}lia, 78735-910, Rondon\'opolis, MT, Brazil}

\email{erikanascimentolima@yahoo.com.br}

\author{T. M. Schmidt}
\affiliation{Instituto de F\'{\i}sica, Universidade Federal de Uberl\^andia,
Caixa Postale 593, 38400-902, Uberl\^andia, MG, Brazil}

\author{R. W. Nunes}
\affiliation{Departamento de F\'{\i}sica, ICEx, Universidade Federal de Minas
Gerais, Avenida Ant\^onio Carlos 6627, Pampulha, 30123-970, Belo
Horizonte, Minas Gerais, Brazil}
\email{rwnunes@fisica.ufmg.br}

\begin{abstract}
We report {\textit ab initio} calculations showing that a single
one-dimensional extended defect can originate topologically-protected
metallic states in the bulk of two-dimensional topological
insulators. We find that a narrow extended defect composed of periodic
units consisting of one octogonal and two pentagonal rings embedded in
the hexagonal bulk of a bismuth bilayer introduces two pairs of
one-dimensional Dirac-fermion states with opposite spin-momentum
locking. Although both Dirac pairs are localized along the
extended-defect core, their interactions are screened due to the
trivial topological nature of the extended defect.
\end{abstract}


\maketitle
Topological insulators (TIs) are a new class of materials
theoretically predicted to exist in 2005~\cite{Kane,Kane2}, with the
first experimental confirmation in HgTe/CdTe quantum wells reported in
2007~\cite{Kane1,Bernevig1,Konig}. TIs have since been subject of
intensive theoretical~\cite{Bernevig1,Konig} and
experimental~\cite{Hsieh} studies due to the coexistence of an
insulating bulk band structure with a non-trivial topology that, when
interfaced with a topologically trivial insulator such as the vacuum,
gives rise to time-reversal-protected metallic surface states, with
Dirac-fermion dispersions spanning the bulk band gap in the
one-dimensional (1D) edges in two-dimensional (2D) TIs. Existence of
the edge states is a requirement imposed by the different topologies
of the band structures across the interface. In these edge states, the
spin quantization axis and the momentum direction are locked-in,
implying that the metallic edge states are protected from
backscattering, rendering their electronic conductance robust against
the presence of disorder. Robust conduction and spin polarization may
allow the manipulation of edge modes of TIs in many applications such
as spintronics~\cite{Garate} and quantum
computation~\cite{Bernevig1,Konig}.
\begin{figure}[h!]
\centering
\includegraphics[scale=0.12,keepaspectratio=true]{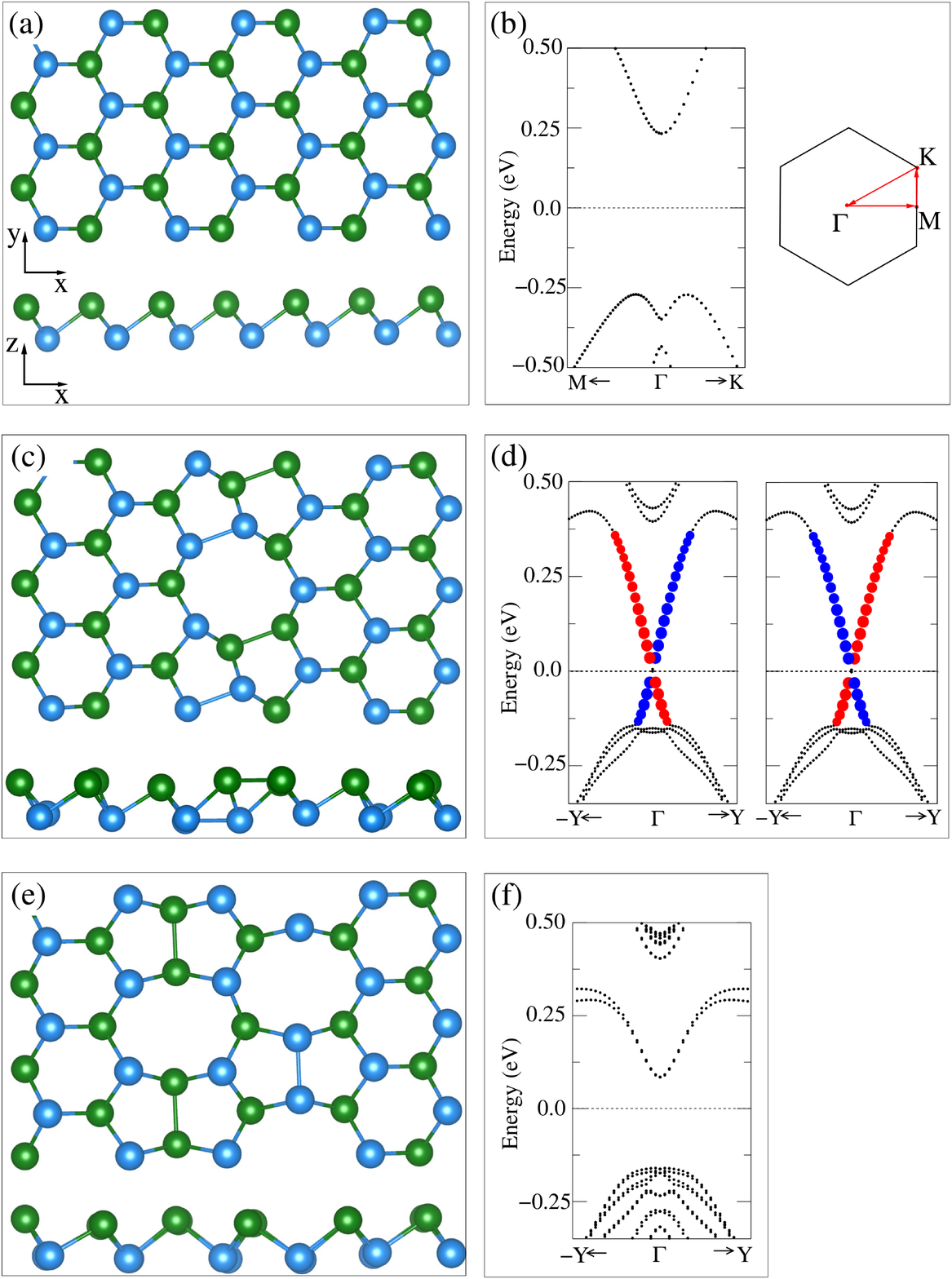}
\caption{(a) Top and side view of a pristine Bi bilayer. (b)
  Electronic band structure of a pristine Bi bilayer. (c) Top and side
  view of a Bi bilayer with an embedded 558 defect (BiBL+558 in the
  text). (d) Band structure of BiBL+558, showing Dirac-fermion
  metallic modes; spin-momentum locking is displayed: blue (red)
  branch of Dirac modes corresponds to spin 'up' ('down'). (e) Top and
  side view of a Bi bilayer with an embedded quadrupole-pentagon
  double-octagon defect (BiBL+Q5D8 in the text); (f) Band structure of
  Bi+Q5D8. In all geometries, atoms in the upper (lower) Bi
  triangular sub-lattice are shown as green (blue) circles.}
\label{fig1}
\end{figure}

While topological-insulating band structures are also found in
three-dimensional (3D) systems, manipulation of the metallic surface
modes in 3D TIs is commonly hampered by the difficulty in tuning the
Fermi level to achieve sufficiently low (ideally null) levels of bulk
carriers, and the metallic surface carriers are significantly
outnumbered by bulk carriers in most 3D TI
samples~\cite{Chen,Checkelsky}. Hence, 2D TIs can be advantageous in
transport applications, because the 2D bulk is fully exposed to
chemical manipulation and, besides, the bulk Fermi level can also be
tuned by proper gating. When an insulating bulk is achieved, electrons
can conduct only along the edge in these structures. From this, many
efforts have been made to find candidate 2D TI systems.

In this work, we show, by means of {\textit ab initio} density functional
theory (DFT) calculations, the formation of ``edge-like'' 1D metallic
electronic states localized on the core of an extended 1D defect
embedded in the bulk of a single (111)-oriented bismuth bilayer
(BiBL). We show that in this system the extended defect implies the
formation of two pairs of topologically-protected Dirac-fermion modes,
each pair localized on each one of the two zigzag edges meeting at the
core of the 1D extended defect. The Dirac pairs are shown to interact
via a strongly screened interaction, and to have opposite chiralities,
which in principle allows for backscattering of carriers propagating
in these defect-induced modes.

Furthermore, we also introduce an alternative crystalline form of a Bi
bilayer, consisting entirely of pentagonal and octogonal rings, which
we refer to as the pentaoctite form of a BiBL. Below, we show that the
pentaoctite BiBL, with a formation energy that is only 0.045~eV/atom
larger than the bulk-derived hexagonal pristine layer, is also a 2D
topological insulator.

Our calculations show that the occurrence of spin-polarized
Dirac-fermion electronic states localized on the core of the extended
line defect, which is fully immersed in the bulk of the 2D TI BiBL, is
closely related to the emergence of metallic states along the edges of
sufficiently large zigzag-terminated bismuth nanoribbons (ZBiNR). The
1D extended defect in our study is a buckled version of the so-called
558-defect, composed of periodic units consisting of one octogonal and
two pentagonal rings, as shown in Fig.\ref{fig1}(c). A flat version of
the 558-defect has been shown experimentally to occur in graphene
monolayers~\cite{558a,558b}, and theoretically to display magnetic
quasi-1D electronic states in n-doped layers~\cite{558c,Lidia}.

While the possible occurrence of metallic fermionic modes along the
core of 1D extended defects (dislocations) in 3D TIs has already been
predicted theoretically for a model tight-binding hamiltonian for a 3D
TI,~\cite{ran} to the best of our knowledge, this work provides the
first demonstration, by {\textit ab initio} calculations, of the emergence
of topologically-dictated helical Dirac-fermion states along the core
of a 1D extended defect immersed in the bulk of a 2D TI.

Structural and electronic properties of extended 1D defects in a BiBL,
in the present work, are computed using the DFT scheme implemented in
the VASP code~\cite{kresse}. The projector-augmented-wave
method~\cite{paw} is used to describe the ionic core-valence electron
interactions. The generalized gradient approximation (GGA) is employed
to describe the exchange and correlation potential.~\cite{gga}
Spin-orbit coupling is included in the calculation of the electronic
structure. Wave functions are expanded in plane waves with energy
cutoff of 300 eV. Convergence with respect to the energy cutoff was
carefully checked from calculations with cuttofs in the 200-400 eV
range. Geometries are optimized until the forces on each atom are less
than 0.03~eV/\AA. Convergence with respect to Brillouin zone (BZ)
sampling was also verfied.

It is well established that a single (111) Bi bilayer in its pristine
form, shown in Fig.\ref{fig1}(a), is a 2D TI (a quantum spin Hall
system)~\cite{Murakami,Wada,Liu,Hirahara1,Yang}. Metallic edge states
are found at the borders of a BiBL finite sheet, with a pair of
helical Dirac fermion states with opposite spin textures (or
spin-momentum locking) on each border~\cite{wang}. The lower panel in
Fig.\ref{fig1}(a) shows that the BiBL is formed from two Bi 2D
triangular sublattices displaced by 1.74~\AA\ along the (111)
direction (the $z$-axis in our supercells). Figure~\ref{fig1}(b) shows
the insulating band structure of an infinite pristine BiBL (thus
devoid of edges) along the $\Gamma$-K and $\Gamma$-M lines in the
Brillouin zone, with a band gap of $\sim$0.5~eV, within the GGA
approximation. The top of the valence band at the $\Gamma$ point is
$\sim$0.1~eV below the absolute maximum of the valence band, so the
direct gap at $\Gamma$ is $\sim$0.6~eV.

The structure of the 558-defect embedded in a (111) BiBL is
illustrated in Fig.~\ref{fig1}(c). The 558-defect is a zigzag-oriented
(along the $y$-axis of the cell) translational grain boundary between
two crystalline domains shifted with respect to each other by
one-third of the lattice period along the armchair direction (the
$x$-axis of the cell). The relative shift leaves a seam between the
two domains that is filled with a roll of Bi dimers, thus forming the
two-pentagon-one-octagon periodic unit of the defect. The period of
the 558-defect is twice the lattice constant of the pristine BiBL
(\textit{a} = 4.33~\AA). In the following, we shall refer to the BiBL
with the embedded 558-defect as BiBL+558. After relaxation, the
BiBL+558 system retains the buckled structure of the pristine
BiBL. The Bi-Bi bond length in a pristine BiBL is 3.04~\AA. In the
pentagonal and octagonal rings of the BiBL+588 bond-lenght values
range from 3.03~\AA\ to 3.11~\AA.

Figure~\ref{fig1}(d) shows the band structure of the supercell for the
BiBL+558 system. Inclusion of the 558-defect leads to the emergence of
two degenerated pairs of edge modes, one on each side of the dimers,
with linear dispersions and the characteristic spin textures of 2D TI
edge modes. Note that the edge-state bands span the band gap of the
insulating bulk BiBL.  Figure~\ref{fig1}(d) shows that the Dirac
dispersions on both sides of the defect have opposite spin textures,
with spin 'up' ('down') states moving 'up' ('down') on the left side
of the dimers and 'down' ('up') on the right side of the
dimers. Hence, the two zigzag lines of Bi atoms at the core of the 558
defect behave as right and left ribbon edges, preserving the
``handedness'' of each edge, which explains the inversion of the spin
texture between the two sides of the Bi dimers.

This is an important point concerning the spin transport of the Dirac
modes along the core of the 558-defect, since, due to the inverted
spin-momentum locking between the Dirac bands on the two sides of the
defect, backscattering is possible by a process where the charge
carriers are scattered to the other side of the defect and propagate
backwards without flipping the spin. The possible formation of 2D
Dirac modes on 2D boundaries (domain or grain boundaries) in 3D TIs
has only been hinted at in Ref.~\onlinecite{ran}, but we anticipate
that the preservation of the handedness of the surface modes may prove
true also in these cases. Hence, from our result for a 1D boundary in
a 2D TI, we expect inversion of the spin texture and the possibility
of backscattering between Dirac modes on the two sides of 2D
boundaries in 3D TIs.
\begin{table}[h]
\centering
\begin{tabular}{|c|c|}
\hline 
$L$ (\AA) & gap (eV) \\ 
\hline                             
12.5 & 0.452 \\
\hline
20.0 & 0.237 \\
\hline 
27.6 & 0.129 \\
\hline
35.1 & 0.071 \\
\hline 
50.7 & 0.026 \\
\hline
58.2 & 0.011 \\
\hline 
99.45 & 0.001 \\
\hline
57.4 (BiBL+558) & 0.022 \\
\hline
77.4 (BiBL+558) & 0.005 \\
\hline 
\end{tabular}
\caption{Energy gap (in eV) for 2D Bi ribbons (ZBiNR) as a
  function of the ribbon width ($L$ in \AA). Last two entries
  show gap for Bi layers with extended 1D defect (BiBL+558) for two
  values of the distance between the 1D defect and its periodic
  images.}
\label{table1}
\end{table}

Below, we address numerically the interaction between the two sets of
Dirac modes on each side of the 558-defect, and come to the conclusion
that the line of dimers at the core of the defect strongly screens the
interaction. We interpret this result as follows: the dimers introduce
a potential barrier that leads to a much faster exponential decay of
the Dirac mode from one side of the defect onto the other side. It is
thus conceivable that backscattering between the two sides of the
defect, as in the above picture, may be inhibited by the same
mechanism.

We propose the following interpretation for the appearance of the
Dirac cones in the BiBL+558 system: the region occupied by the
558-defect behaves topologically as a trivial insulating material,
akin to the vacuum surrounding the edges of a bismuth ribbon.
Furthermore, it has been determined by Wang {\textit et al.}~\cite{wang}
that by saturating the edges of a ZBiNR with H atoms, the Dirac cones
of topological edge states move from the K point in the case of
unsaturated edges to the $\Gamma$ point in the H-saturated-edge
case. We observe the same behavior in the metallic states along the
core of the 558-defect in a BiBL: the Dirac point for the pair of
degenderate Dirac branches occurs at the $\Gamma$ point. Hence, the
line of Bi dimers in the core of the defect plays a role of saturation
likewise that of hydrogens in a ZBiNR.

We will expand on the above interpretation in the following, but let
us first analyze the nature of the interaction between the two pairs
of Dirac modes on the core of the 558-defect. 

\begin{figure}
\centering
\includegraphics[scale=0.4,keepaspectratio=true]{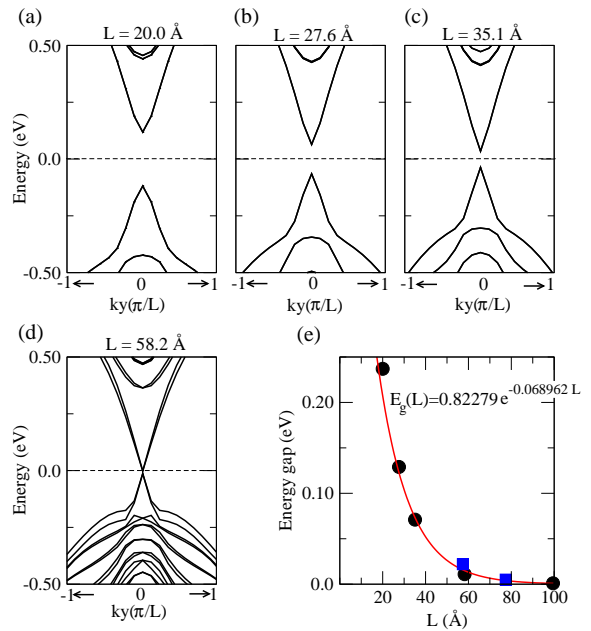}
\caption{(a)-(d) Electronic band structures of zigzag-edged Bi
ribbons (ZBiNR), with widths ($L$) ranging from 20 to 58~\AA; (e)
the band gap evolution of ZBiNRs as a function of $L$. Exponential
fitting for gap$\times~\;L$ is shown. Blue squares show gap values for
Bi layers with embedded 1D extended defect for two values of the
distance between the 1D defect in the cell and its periodic images.}
\label{fig2}
\end{figure}

We start from the behavior of the gap of the Dirac edge modes of a
ZBiNR, due to the quantum tunneling between the topological states
with the same spin alignment in the two edges of the ribbon. The gap
in the Dirac-fermion dispersion at the edges of the ZBiNR reflects the
fact that the edge states penetrate into the bulk, with an exponential
decay, and the interaction leads to the opening of a
bonding-antibonding gap. As shown in Fig.~\ref{fig2} and
Table~\ref{table1}, as we increase the width of the ribbon the gap
decreases, as expected. Figure~\ref{fig2}(e), shows an exponential
fitting of the ribbon gap as a function of the ribbon width ($L$). For
a $\sim$58~\AA -wide ribbon the gap is 11~meV.

For the BiBL+558 system, due to the periodic boundary conditions of
the supercell, a similar interaction due to the quantum tunneling is
expected between the Dirac modes of the defect in the supercell and
their neighboring periodic images. In the case of the ribbons, each
edge interacts only with the other edge, while in the BiBL+558 the
Dirac modes along the 558-defect core interact primarily with their
two nearest periodic images, one to the right and the other to the
left of the defect in the cell. The last two entries in
Table~\ref{table1} show the gap of the edge modes for the BiBL+558,
for two different widths of the pristine region of the supercell. The
gap for the supercell with a $\sim$57~\AA-wide pristine region is
22~meV, twice the value for a ribbon this wide as included in
Table~\ref{table1}, which indicates that the interactions with the two
periodic images add up.

The above result is also suggestive of a very weak interaction between
the two sets of Dirac modes in the core of the 558-defect, mediated by
the line of dimers in the center of the core. In the BiBL+558, the
distance between the edges on each side of the dimers is 5.35~\AA. In
order to investigate the nature of this interaction, we have also
computed the gap in the Dirac modes for two H-saturated ribbons
interacting through the vacuum in the same geometric configuration as
in the BiBL+558 system, i.e., we removed the line of Bi dimers and
kept the distance between the edges of the ribbons unchanged (at
5.35~\AA). In this case, the gap increased by an order of magnitude to
134~meV. This result indicates that in the BiBL+558 the line of dimers
strongly screens the interaction between the Dirac edge states. This
is further confirmed by a calculation of the gap for a
$\sim$77~\AA-wide BiBL+558 system that yields a gap that falls on top
of the exponential fitting for an isolated H-saturated ribbon, as
shown by the red square in Fig.~\ref{fig2}.

\begin{figure}
\centering
\includegraphics[scale=0.15,keepaspectratio=true]{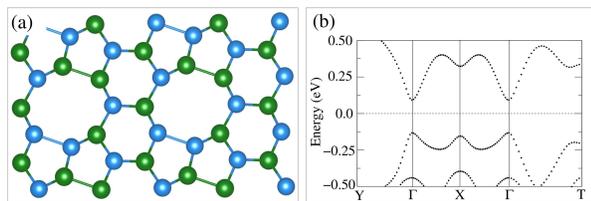}
\caption{(a) and (b) show a top of view and the band structure of a
penta-hexa-octo form of a 2D Bi layer, respectively.}
\label{fig3}
\end{figure}

Let us go back now to our interpretation for the emergence of
topological edge states along the core of the 558-defect. It rests on
the assumption that the defect region acts as a portion of a trivial
insulator. In order to confirm this picture, we compute the $Z_2$
topological invariant for a pristine BiBL and for two ``defective'' Bi
2D layers. For a pristine BiBL, the $Z_2$ has been obtained by looking
at the parity of the occupied bands at the time-reversal-invariant
momenta (TRIMs). We obtain $Z_2 = -1$, confirming its non-trivial
topological character, with a band inversion at the M point. For a 2D
bulk composed by a sequence of 558-defects connected by a stripe of
single hexagons, as shown in Fig.~\ref{fig3}(a), from the parity of
the bands at the TRIMs we obtain $Z_2 = 1$, showing that, indeed, this
stripe of a hypothetical penta-hexa-octo form of 2D Bi bilayers is a
trivial insulator. The band structure for the penta-hexa-octo Bi
bilayer is shown in Fig.~\ref{fig3}(b).

A corollary of the above reasoning is that if we insert a stripe of a
topological insulator as an extended 1D defect in a BiBL, we would see
no Dirac cones of topologically-protected ``edge'' states along such a
defect. In order to test this idea, we insert another extended defect,
the quadruple pentagon-double octagon (Q5D8) defect, shown in
Fig.~\ref{fig1}(e), inside the BiBL. The Q5D8-defect is formed by two
adjacent stripes of the 558-defect. As shown in Fig.~\ref{fig1}(f),
the Q5D8-defect does not introduce Dirac modes in the band gap of the
BiBL, and a band gap is always present in the electronic structure of
the BiBL+Q5D8 supercell. It follows that the Q5D8-defect can be seen
as a finite portion of a Bi structure that belongs to the same
non-trivial topological class as the pristine BiBL.

Although the 558-defect and the Q5D8 are both formed by pentagons and
octagons, there is a crucial structural difference between the two
defects. A pristine BiBL consists of two Bi triangular sublattices,
labeled A and B, which are shiftted by 1.74~\AA\ in the perpendicular
direction. In this geometry, each Bi atom in the ``upper'' (B)
sublattice is bonded to three Bi atoms in the ``lower'' (A)
sublattice, and vice-versa, as shown in Fig.~\ref{fig1}(a). When we
insert the Q5D8-defect in a pristine BiBL, all Bi-Bi bonds along the
direction perpendicular to the defect are the same as in the pristine
BiBL [a sequence of Bi$_{\rm B}$-Bi$_{\rm A}$ bonds, see Fig.~\ref{fig1}(e)],
and this system belongs to the same topological class as the BiBL
itself. On the other hand, the BiBL+558 system [see Fig.~\ref{fig1}(c)]
presents Bi$_{\rm B}$-Bi$_{\rm B}$ and Bi$_{\rm A}$-Bi$_{\rm A}$
bonds, which brake the A-B sublattice bonding pattern, and as a
consequence it does not present a band inversion, and thus belongs to
the class of trivial of insulators.

\begin{figure}
\centering
\includegraphics[scale=0.32,keepaspectratio=true]{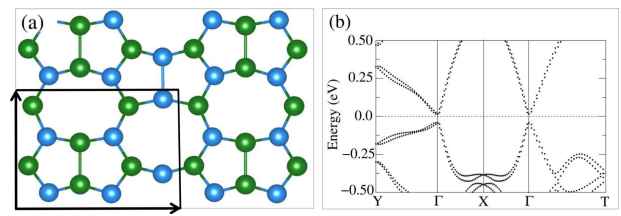}
\caption{(a) and (b) show a top of view and the band structure of a
pentaoctite form of a 2D Bi layer, respectively.}
\label{fig4}
\end{figure}

Our result for the electronic structure of the BiBL+Q5D8 system,
suggests a striking conclusion, that an extended crystaline form of
the Q5D8-defect should be a topological insulator.  Based on that
observation, we decided to investigate the topology of the electronic
structure of a pentaoctite form of a BiBL, as shown in
Fig.~\ref{fig4}(a). In this geometry, the bilayer consists of buckled
pentagons and octagons only. Figure~\ref{fig4}(a) shows a 2x2 cell of
the pentaoctite BiBL. The rectangular box shows the 12-atom unit cell
for this periodic structure. Indeed, this pentaoctite BiBL is an
isulator, with a small gap of 0.035~eV, at the GGA level. From a
calculation employing a hybrid functional we obtained a robust gap of
0.45~eV. For the calculation of the $Z_2$ topological invariant of the
Bi pentaoctite, we found it more expedient to employ the real-space
methodology introduced by Soluyanov and Vanderbilt~\cite{dhv}. We
obtain $Z_2$=-1, confirming our expectation that the pentaoctite BiBL
is a 2D topological insulator.

The fact that a hypotetical pentaoctite form of a Bi bilayer is a
topological insulator naturally leads us to examine its possible
experimental realization. In Table~\ref{table2}, we show a comparison
of the formation energy and atomic density of the bulk-derived
hexagonal BiBL and of the pentaoctite BiBL. The energy of the
hexagonal layer is set to zero, for reference.  The formation energy
of the pentaoctite BiBL is only 0.045~eV/atom above that of the
hexagonal BiBL, i.e., the energy difference between the two phases is
of the order of the thermal energy fluctuations per atom, at room
temperature. Thus, from the energetics point of view the pentaoctite
BiBL is a viable crystalline phase. Table~\ref{table2} also shows the
atomic density of the two phases. The pentaoctite BiBL is a less dense
phase than the hexagonal layer, which indicates that a BiBL may undergo
a phase transition to the pentaocite form under a tensile strain. The
pathway for a hexagonal-BiBL~$\rightarrow$~pentaoctite-BiBL transition
will be the subject of a forthcoming publication.
\begin{table}[h]
\centering
\begin{tabular}{|c|c|c|}
\hline
structure & formation energy & atomic density\\ 
\hline 
pristine & 0.000    & 0.1231 \\
\hline
pentaoctite & 0.045  & 0.1174 \\
\hline
\end{tabular}
\caption{Formation energy (in eV/atom) and atomic density (in
  atom/\AA$^2$) of a bulk-derived hexagonal bismuth bilayer (BiBL) and
  of a pentaoctite form of the BiBL. Formation energy of the hexagonal
  layer is set to zero.}
 \label{table2}
\end{table}

In conclusion, {\textit ab initio} DFT calculations indicate the formation
of helical Dirac-fermion metallic states localized on the core of an
extended 1D defect embedded in the bulk of a single (111)-oriented
bismuth bilayer (BiBL). The formation of topologically-protected
metallic states on the core of the 1D defect is related to the
emergence of such metallic states along the edges of sufficiently
large Bi ribbons. The core of the extended defect in our study is
composed of two zigzag chains of Bi atoms connected by a roll of Bi
dimers, forming the periodic unit of the defect, consisting of one
octogonal and two pentagonal rings. The two zigzag lines of Bi atoms
on the core of the 1D defects act as two zigzag egdes, each edge
hosting a pair of Dirac-fermion metallic modes, with linear
dispersions.

Numerical DFT results show that the interaction between the Dirac
modes on the two sides of the defect is strongly screened by the roll
of Bi dimers at the geometric center of the 1D defect core.  The
handedness of the two zigzag ``edges'' meeting at the core of the 1D
defect is preserved, which leads to an inversion of the spin-momentum
locking between the two pair of Dirac modes localized on the 1D
defect: spin 'up' ('down') modes propagate up (down) on one side of
the defect and down (up) on the other side, which leads to the
possibility of backscattering between the Dirac modes, induced by
disorder in the region of the 1D extended-defect core.

Moreover, we also introduce an alternative crystalline form of a Bi
bilayer, consisting entirely of pentagonal and octogonal rings, which
we refer to as the pentaoctite form of a Bi bilayer. The formation
energy of the pentaoctite Bi bilayer is only 0.045~eV/atom larger than
that of the bulk-derived hexagonal layer, and its density is slightly
larger than that of the hexagonal layer, indicating a possible
structural transition between the two 2D crystalline phases at
moderate imposed tensile biaxial strains. Computation of the $Z_2$
topological invariant shows that the pentaoctite BiBL is a 2D
topological insulator.

\bibliography{bi558}

\end{document}